\documentclass[journal]{IEEEtran}

\usepackage{graphicx}
\usepackage{epstopdf}
\usepackage{amsmath}
\usepackage{amssymb}
\usepackage{amsthm }
\usepackage{epsfig}
\usepackage{times}
\usepackage{setspace}
\usepackage{bm}
\usepackage{siunitx}
\usepackage{threeparttable}
\usepackage{balance}
\usepackage{tabularx}
\usepackage{subcaption}
\usepackage{xcolor}
\usepackage{fancyhdr}

\usepackage{array}

\usepackage[T1]{fontenc}
\usepackage{times}
\usepackage[mathscr]{eucal}
\usepackage{wrapfig}
\usepackage{multirow}
\usepackage{tablefootnote}

\newcommand{\equals}{\!=\!}

\newcommand{\minus}{\!-\!}
\newcommand{\gteq}{\!\ge\!}
\newcommand{\gthan}{\!>\!}
\newcommand{\lthan}{\!<\!}
\newcommand{\plus}{\!+\!}

\newcommand{\e}{\mathrm{e}}

\newcommand{\SF}{\mathrm{SF}}

\begin{document}

\title{Accelerating Update Broadcasts Over LoRaWAN Downlink via D2D Cooperation}

\author{Anshika Singh and Siddhartha S. Borkotoky, \IEEEmembership{Member, IEEE} 
\thanks{
This work was supported by the Department of Science and Technology (DST), Government of India, through its Core Research Grant (CRG) No. CRG/2023/005803. The authors are with the School of Electrical and Computer Sciences, Indian Institute of Technology Bhubaneswar, Khordha 752050, India (e-mail:a25ec09003@iitbbs.ac.in, borkotoky@iitbbs.ac.in)}
} 
\maketitle

\thispagestyle{fancy}
\lhead{{\color{gray} This manuscript has been accepted for publication in the \textit{IEEE Transactions on Industrial Informatics}. This is the author's version, and is posted here for personal use, not for redistribution.}}

\begin{abstract}
Broadcast distribution of updates (e.g., security patches, machine learning models) from a server to end devices (EDs) is a critical requirement in the Internet of Things (IoT). In this paper, we consider the problem of reliable over-the-air broadcast of updates in Long Range Wide Area Networks (LoRaWANs). Existing broadcast techniques for LoRaWANs suffer from long delivery delays due to low data rates and duty-cycle constraints. We address this problem by proposing a device-level cooperative mechanism,
in which updated EDs broadcast a few update fragments to accelerate delivery to their neighbors. We demonstrate large reductions in the delivery time compared to conventional methods. For instance, in a 400-node network spanning 1 km radius and operating at 1\% duty-cycle, the proposed scheme reduces the time required to deliver a 10 kilobyte update to an ED at the network’s edge from 42 hours to 45 minutes. The proposed solution thus provides a pathway toward improved security and efficient realization of edge intelligence in LoRaWAN IoT. 
\end{abstract}


\section{Introduction}
\label{sec:intro}
The LoRaWAN communication standard is characterized by its ability to support low-power, long-range data transmission. It has thus enabled a broad range of IoT applications such as smart cities, smart agriculture, remote monitoring, and Industry 4.0. In such IoT deployments, it is often necessary to broadcast updates to end devices (EDs). These updates may include security patches that address newly discovered vulnerabilities or firmware upgrades that introduce new functionalities. Update broadcasts are also essential for supporting distributed machine learning (ML) in IoT networks. For instance, in the popular distributed-ML paradigm known as federated learning, a server must iteratively broadcast global ML model updates to EDs.

\begin{figure}
    \centering
    \includegraphics[scale=0.32]{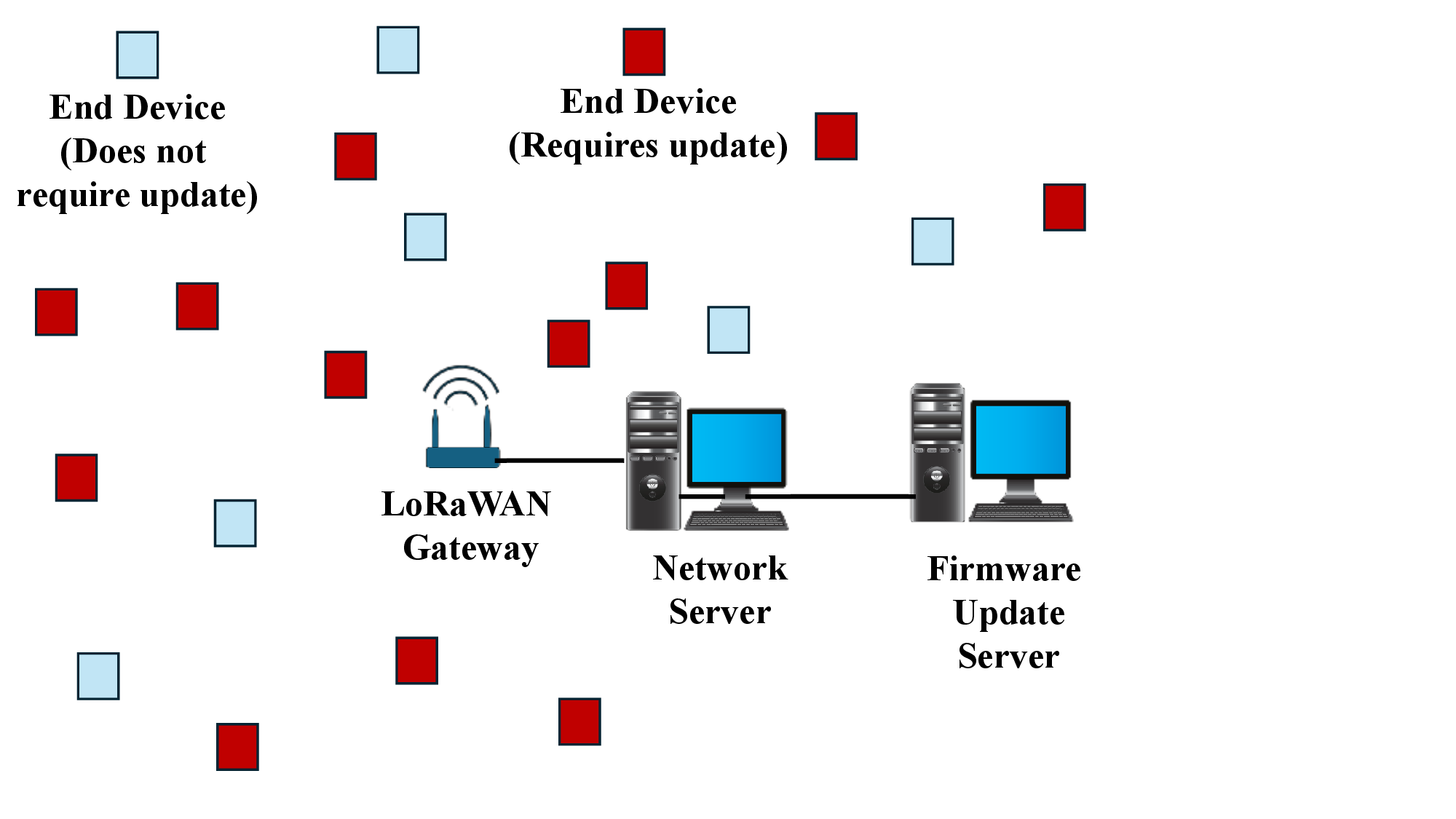}\caption{Representative model of a LoRaWAN in which a subset of the EDs require a firmware update. The update recipients are shown in red. The gateway multicasts the update frames produced by the firmware update server.}\
    \vspace{-8mm}
    \label{fig:network_model}
\end{figure}

The need for reliable update dissemination has led to the development of the paradigm called firmware updates over the air (FUOTA). The FUOTA Working Group within the LoRa Alliance defines guidelines for FUOTA over LoRaWAN~\cite{FUOTA}. In this framework, an update server divides the update into fragments, each small enough to fit a LoRaWAN link-layer frame. The frames are multicast via a gateway to all EDs that require the update. A representative FUOTA model for LoRaWAN is shown in Fig.~\ref{fig:network_model}. The FUOTA framework can also be used to transfer other updates, such as an ML model. 
Therefore, in the remainder of this paper, we use FUOTA as an umbrella term for the broadcast or multicast transfer of updates (firmware, ML model, etc.) from a gateway to multiple EDs in a LoRaWAN. 

A key challenge in FUOTA is the long time required to deliver updates. LoRaWAN frames have modest data rates, sometimes as low as 300 bps. The transmissions are further constrained by strict regional duty-cycle limits. For instance, devices in the 868 MHz band in the EU are limited to a 1\% transmitter duty cycle, necessitating idle periods between successive update fragments. These limitations can lead to substantial delivery delays. For example, consider a 10 kB update divided into 200 fragments of 50 bytes each. With spreading factor (SF) 12 and 125 kHz bandwidth, each fragment has an on-air time of roughly 2.8 s, yielding a total of 560 s of airtime. Under a 1\% duty cycle, the minimum delivery time becomes 56,000 s ($\approx 15$ h). In practice, some fragments will be lost and retransmitted, leading to a much longer delivery time. This can be unacceptable for time-sensitive security patches and can significantly delay the convergence of iterative distributed ML methods such as federated learning.

Using higher data rates (lower SFs) can reduce update time, but only when the frame-loss rate is low. To ensure acceptable reliability for all recipients including those far from the gateway, low data rates (high SFs) are typically required. A practical workaround is to deploy multiple gateways. Each gateway then covers a smaller area and can transmit at higher data rates. However, this raises infrastructure cost and undermines LoRaWAN’s primary advantage of wide-area coverage.

As a remedy, we propose a method to accelerate update delivery in LoRaWAN without deploying additional gateways. The approach leverages device-to-device (D2D) communication: once an ED receives the complete update, it broadcasts a few fragments to other EDs. These unfinished EDs then collect fragments from both the D2D links and the gateway, yielding significantly faster update completion than conventional gateway-only delivery.

\section{Related Work}
The FUOTA working group has released multiple specifications addressing different aspects of FUOTA~\cite{FUOTA_MCAST, FUOTA_SYNC, FUOTA_FRAG, FUOTA_FMP, FUOTA_MPA}. Of particular interest is~\cite{FUOTA_FRAG}, which considers the fragmentation of the update for transmission as link-layer frames. Noting the need to mitigate the effect of frame erasures for reliable update delivery,~\cite{FUOTA_FRAG} recommends the use of forward-erasure correction (FEC). In FEC, the source transmits redundant (coded) fragments, which are linear combinations of source fragments. The redundancy helps compensate for frame losses, thus improving reliability.


In~\cite{AFM20}, the authors study the effect of link-layer parameters on FUOTA performance. FEC is used for frame recovery. They report that lower data rates improve update efficiency but increase delivery delay, consistent with the findings of~\cite{ACC20}. Practical FUOTA-based dissemination of tiny-ML models in LoRaWAN is demonstrated in~\cite{NNA22}, where a pre-trained model is broadcast to EDs deployed on an agricultural field. 

The above approaches employ a best-effort model. In this model, the update session ends after a fixed number of fragments have been sent. However, this does not ensure the delivery of the complete update to all EDs. Rateless-coding–based FEC solutions in~\cite{SNY23,SNY24} address this by allowing the gateway to continue transmitting coded fragments until all devices complete the update. In the multi-SF group-based method of~\cite{MZK23}, EDs are grouped by distance and assigned distinct SFs, with multicast performed group-wise; this reduces energy consumption but increases delay~\cite{Bor25}. A tunable delay–energy trade-off is provided in~\cite{Bor25}, where the gateway cycles through SFs without grouping EDs.

All these methods rely solely on gateway transmissions. To our knowledge, this paper is the first to consider a D2D-aided strategy for FUOTA over LoRaWAN. Indeed, D2D communications have been studied in many contexts, ranging from industrial IoT~\cite{OzP23} to cognitive radio networks~\cite{GMC24}. Examples of the use of D2D communications in LoRaWANs include~\cite{MPH17,dLR22,SSB22}. However, existing works either consider the establishment of D2D data transfer sessions with the help of the network server~\cite{MPH17} or use EDs as D2D relays to transmit device data on the uplink~\cite{dLR22,SSB22}. We instead target reliable downlink delivery of complete updates using scalable, server-independent D2D communications among EDs.

Rateless codes lend themselves well to device-level cooperation for accelerating update transfer. This is illustrated, for example, in~\cite{BoP19,VRF10,BoP15}. In these works, broadcast recipients assist the source by forwarding partial updates to unfinished neighbors. However, these approaches assume comparatively capable devices. Such devices can perform neighbor probing, establish point-to-point or multipoint links, create feedback channels, and continue transmissions until all neighbors complete the update. These mechanisms are impractical for low-power IoT networks with unsophisticated EDs that face stringent energy, bandwidth, and duty-cycle limitations. Our method is designed for such networks. It does not require neighbor discovery, D2D transmitter–receiver association, and topology knowledge.

\section{Background}
\subsection{LoRaWAN}
LoRaWAN is a link-layer protocol that supports low-power wide area networks. The proprietary physical layer communication technology used in a LoRaWAN is referred to as LoRa, which is a form of chirp spread spectrum modulation. A LoRaWAN frame contains a preamble for synchronization and acquisition, an optional header, and the payload. The SF determines the airtime of a frame. SF values are integers ranging from 7 through 12. The duration of a frame with SF $i$ and a payload of $b$ bytes is given by
\begin{align} \nonumber
    l_i(b) =& \bigg[n_{\mathrm{pr}} + 12.5 \max\left\{ \left\lceil \frac{2b - i - 5h + 11}{i - 2y} \right\rceil (c \plus 4), 0\right\}\bigg]\\ 
    &\times {2^{i}}/{\mathrm{BW}}, 
\end{align}
where $n_{\mathrm{pr}}$ is the number of preamble symbols, $\mathrm{BW}$ is the transmission bandwidth, $h$ equals 1 if the frame includes a header and 0 otherwise, $y$ equals 1 when low data rate optimization is enabled and 0 otherwise, and $c$ is an integer from 1 to 4 depending on the channel code applied to the payload~\cite{Sem13}. Increasing the SF increases the frame duration, when other parameters ae fixed. The SF also dictates the receiver sensitivity. Higher SFs provide better sensitivity, enabling reception of lower-power signals. 

LoRaWANs by default follow a star topology, in which EDs communicate directly with a central gateway. Uplink transmissions typically use pure ALOHA to avoid scheduling complexity. For downlink communication, EDs can operate in one of three classes. The most widely used and energy-efficient is Class A, in which the ED opens two receive windows after sending an uplink message. In Class B, the EDs open additional receive windows (ping slots) at scheduled intervals, using gateway-sent beacons for timing. Class C devices continuously listen for incoming messages, except during their own transmissions.

\vspace{-2mm}

\subsection{Summary of FUOTA}
As specified by the FUOTA working group, the FUOTA process can be divided in several subtasks, including multicast-group formation~\cite{FUOTA_MCAST}, device synchronization~\cite{FUOTA_SYNC}, link-layer update fragmentation~\cite{FUOTA_FRAG}, firmware management~\cite{FUOTA_FMP}, and multi-package access~\cite{FUOTA_MPA}. An in-depth account of the FUOTA specifications released by the LoRa Alliance's FUOTA working group can be found in~\cite{AFM20} and~\cite{FUOTAFAQ}. 

The \textit{firmware update server} and the \textit{firmware update management entity} are responsible for managing the FUOTA process. The two entities coordinate with a network server to transmit the firmware update to the EDs. 
Due to the relatively large size of the update, it is divided into smaller \textit{fragments}~\cite{FUOTA_FRAG}. Each fragment is sent as the payload of a single LoRa frame. Fragment transmission is preceded by a setup phase in which the server creates a multicast group of the update recipients. Upon group formation, the gateway begins the multicast transmissions. For successful reception, EDs must know the gateway’s transmission times. The necessary synchronization procedure toward this end is included in the FUOTA specifications~\cite{FUOTA_SYNC}. 

Note that Class-A receptions are not suitable for FUOTA. EDs operating in this class can listen only for short time following an uplink transmission, whereas FUOTA requires a sequence of receptions on the downlink.  Consequently, FUOTA specifications require Class-A EDs to switch to Class B or C for the duration of the update session. 

\vspace{-1mm}
\subsection{FEC}
\label{sec:FEC}
Update completion at an ED requires the successful delivery of all source fragments. This necessitates a mechanism to compensate for frame losses. FEC provides a scalable solution to this problem as follows: Suppose the update comprises $k$ source fragments. The transmitter transmits \textit{coded fragments}, which are linear combinations (bitwise XOR) of certain source fragments. Consider an ED attempting to receive a stream of such coded fragments. As soon as the ED receives \textit{any} set of coded fragments containing $k$ linearly independent combinations of the source fragments, it can decode the update (i.e., reconstruct all original source fragments) by solving a system of linear equations via Gaussian elimination or other specialized decoding algorithms.

The primary benefit of FEC is that the gateway need not track which transmissions have been received by which ED or retransmit lost fragments. Consequently, the EDs do not need to acknowledge individual frame receptions. Instead, an ED sends a single acknowledgment to the gateway upon successfully decoding the update.

The procedure for selecting source fragments to linearly combine into a coded fragment depends on the type of FEC used. The FUOTA specifications include an illustrative example in which the source fragments are chosen according to an LDPC encoding matrix~\cite{FUOTA_FRAG}. The number of coded fragments to be transmitted (say, $k'$) is greater than $k$ and is fixed \textit{a priori}. The ratio $k/k'$ is called the code rate.

\subsection{Rateless FEC}
In conventional FEC, the EDs that receive fewer than $k$ linearly independent coded fragments out of the $k'$ transmissions fail to reconstruct the update. Rateless codes eliminate this limitation by generating a potentially infinite stream of coded fragments. This permits the source to continue transmissions until all EDs decode the update.

A rateless-coded fragment is the sum of pseudorandomly selected source fragments, generated as follows: First, an integer $D$ is chosen at random from the range $[1,k]$ according to a probability mass function called the \textit{degree distribution}. Next, $D$  source fragments are randomly selected and summed using bitwise XOR  to produce the coded fragment.  

For decoding, the receiver must know the sequence numbers of the source fragments chosen for each coded fragment. While this information can be included in the frame header, it increases the communication load. Instead, practical rateless codes employ a predefined mapping (produced using a known pseudorandom pattern)  between the sequence number of the coded fragment and the constituent source fragments. Hence, including only the coded fragment's sequence number in the frame allows the receiver to infer the necessary information.

Well-known practical rateless codes include LT codes~\cite{Lub02}, raptor codes~\cite{Sho06, BoP19}, and random linear network codes~\cite{BoP15,HST08}. 
Note that FEC incurs computational overheads at the source and destination. However, significant progress has been made toward developing rateless codes suitable for resource-constrained scenarios. Examples include a TinyOS implementation of a random linear code for sensor networks~\cite{HST08} and   
a rateless code designed for FUOTA over LoRaWAN~\cite{SNY23}. The availability of these methods, coupled with the LoRa Alliance's recommendation to employ FEC for FUOTA~\cite{FUOTA_FRAG}, leads us to believe that FEC is viable for reliable data transfer in LoRaWANs.

\section{Proposed Scheme}
\label{sec:proposed_scheme}
The following description assumes that the FUOTA session configuration phase has been successfully completed. That is, the multicast group of update recipients has been created and the initial control signaling is complete.

\subsection{Basic Structure of the Protocol} 
The update is divided into $k$ source fragments of $b$ bytes each. The gateway applies rateless coding to the source fragments to produce coded fragments, and transmits these as a stream of \textit{downlink frames}. Each ED attempts to receive all transmissions until it decodes the update. Upon successful decoding, an ED broadcasts a few \textit{D2D frames}, each carrying a coded fragment generated locally at the sender. The number of D2D frames is determined  adaptively, using the procedure described in Section~\ref{sec:N2D}. The D2D frames have two purposes. First, they serve as an acknowledgment of update completion. Upon receiving a D2D frame, the gateway marks the sender as updated. Secondly, unfinished EDs listen for incoming D2D frames. If a D2D frame is received successfully, the recipient adds the encapsulated coded fragment to its set of received fragments. This allows recipients to accumulate coded fragments faster than with gateway-only reception and expedites update delivery. 

In terms of selecting frequency bands and SFs for the frames, the proposed method can be integrated with any physical-layer parameter-selection mechanism. For simplicity, here we assume that all transmissions (downlink and D2D) occur over the same frequency band, which is announced during the session set-up phase.  To assign SFs for downlink frames, we employ the method of~\cite{Bor25}. In this method, multiple SFs are used sequentially. The gateway starts with SF $L$, where $7 \leq L \leq 12$. After every $w$ frame transmission (where $w$ is a design parameter), the SF is incremented by one. There is no further increase once SF 12 is reached. This approach enables the EDs with strong links from the gateway to receive the update early (since lower SF frames are shorter), thus expediting the start of the D2D data flow. The D2D transmissions use a fixed SF, referred to as $\mathrm{SF}_{\mathrm{D2D}}$. 

The implementation details are described next. The important notations are summarized in Table~\ref{tab:notation}.

\begin{table}[]
    \centering
    \caption{Summary of key notation}
    \begin{tabular}{c|c}
        Sym. &  Definition \\ \hline
        $k$ & No. of fragments in the update\\
        $b$ & Bytes per fragment\\
        $l_i(b)$ & Frame duration for SF $i$ and $b$ bytes payload\\
        $N_{\mathrm{ED}}$ & No. of EDs to be updated\\
        $\tau$ & Percentage duty cycle\\
        $T_p$ & Ping-slot (PS) duration\\
        $G_p$ & No. of PSs occupied by a downlink frame\\
        $E_p$ & No. of PSs occupied by a D2D frame\\
        $W_p$ & No. of PSs between two consecutive DL frames\\
        $S_{\mathrm{D2D}}$ & No. of superslots in a D2D window\\
        $S^*$ & Largest permitted value of $S_{\mathrm{D2D}}$\\
        $N^{(\max)}_{\mathrm{D2D}}$ & Maximum possible D2D transmissions by an ED\\
        $N^{(\min)}_{\mathrm{D2D}}$ & Minimum possible D2D transmissions by an ED\\
        $\SF_{\mathrm{D2D}}$ & SF used for D2D frames\\
        $M$ & Max. no. of frames that can be sent by gateway\\
        $L$ & SF with which gateway (GW) begins transmissions\\
        $w$ & No. of frames after which GW increments the SF
    \end{tabular}
    \label{tab:notation}
\end{table}

\subsection{Downlink Transmission Schedule}
\label{sec:DL_schedule}
\begin{figure}
    \centering
    \includegraphics[scale=0.28,bb=350 110 620 480]{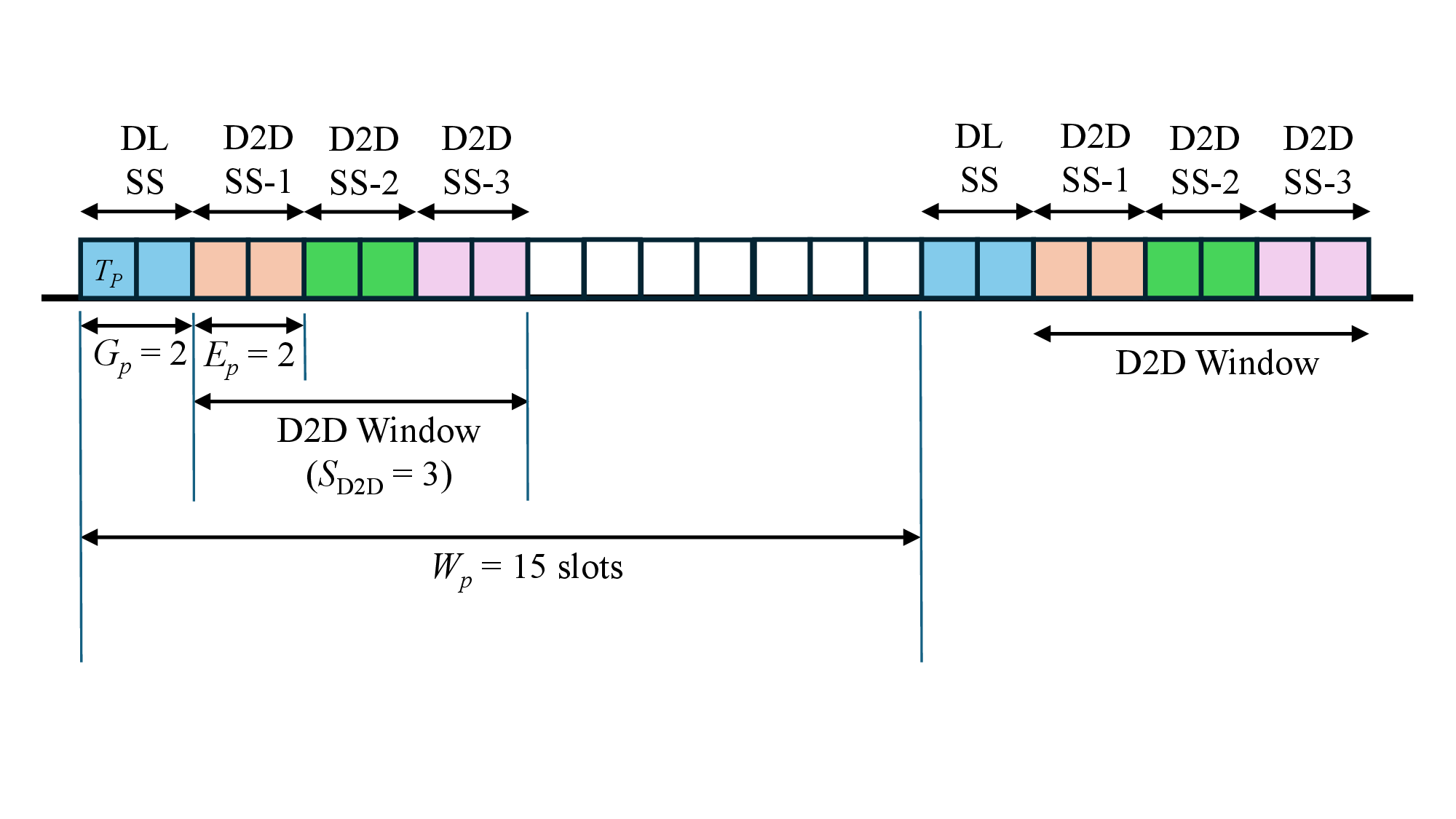}
    \caption{Illustration of downlink (DL) and D2D transmission schedules. Each  block represents a ping slot of duration $T_p$. The schedule is illustrated for the case in which one superslot (SS) for both DL and D2D links is composed of two ping slots ($G_p \equals 2$, $E_p \equals 2$), one D2D window comprises 3 D2D superslots ($S_\mathrm{D2D} \equals 3$), and 15 ping slots must elapse between the starts of two consecutive DL frames ($W_p \equals 15$) to satisfy duty-cycle restrictions. }
    \label{fig:tx_schedule}
\end{figure}

Our proposed scheme can be used with both Class B and Class C LoRaWAN. Here we assume Class B owing to its superior energy efficiency. Recall that in Class B, time is divided into ping slots. The transmission of a downlink frame from the gateway starts at the beginning of a ping slot. Depending on the frame duration, the transmission may occupy a non-integer number of slots. We denote the collection of slots (full and partial) occupied by a frame transmission as one \textit{downlink superslot}. 
Therefore, the number of ping slots in a downlink superslot is 
\begin{align}
    G_p = \left\lceil\frac{l_i(b)}{T_p}\right\rceil,
\end{align}
where $i$ is the SF used, $b$ is the fragment size in bytes, $T_p$ is the ping-slot duration, and $\lceil x \rceil$ denotes the smallest integer greater than or equal to $x$. An illustration of the case for $G_p \equals 2$ is shown in Fig.~\ref{fig:tx_schedule}.

To satisfy the duty-cycle restrictions, the gateway  enters a silent period following every downlink frame. The number of slots that must elapse between consecutive downlink transmissions is computed as follows: Consider a downlink frame with SF $i$ and duration $l_i(b)$. To maintain a duty cycle of $\tau$\% in the absence of time slotting, the time $W_t$ from the start of the said transmission to the start of the next must satisfy
\begin{align}
    \frac{l_i(b)}{W_t} = \frac{ \tau}{100}. 
\end{align}
Since time is divided into ping slots, the number of slots elapsed since the beginning of the downlink transmission until the start of the next is
\begin{align}
\label{eq:Wp}
    W_p = \left\lceil \frac{W_t}{T_p}\right\rceil = \left\lceil \frac{100 \times l_i(b)}{  \tau \times T_p }\right\rceil.
\end{align}

The  sequence numbers of the coded fragments (see Section~\ref{sec:FEC}) sent by the gateway start from 0 and are incremented by one after every downlink transmission. That is, the coded fragment encapsulated in the first frame from the gateway has  sequence number 0, the second frame has sequence number 1, and so on. The gateway transmits a maximum of $M$ coded fragments. Hence, the largest possible sequence number that can be used by the gateway is $M \minus 1$.

We remark here that restricting the maximum number of transmissions to $M$ seemingly violates the rateless-coding principle of transmitting a potentially infinite stream of redundancy. However, for all practical purposes, using a large $M$ ensures that the code is virtually rateless. For example, in our numerical results, $K \equals 200$ and $M \equals 10,000$, permitting the transmission of up to 9800 redundant fragments.   

\subsection{The D2D Window}
\label{sec:D2D_window}
Each ED transmits D2D frames after receiving the update. Let $E_p$ be the number of ping slots occupied by a D2D frame carrying a coded fragment from an ED. Observe that
\begin{align}
\label{eq:Ep}
    E_p = \left\lceil\frac{l_j(b)}{T_p}\right\rceil,
\end{align}
where $j \equals \SF_{\mathrm{D2D}}$. A \textit{D2D window} comprising $S_\mathrm{D2D}$ superslots is reserved after a downlink superslot. 
This is illustrated in Fig.~\ref{fig:tx_schedule}. To prevent the D2D window from encroaching into the next downlink superslot, 
we must have
\begin{align}
\label{eq:SD2D_bound}
    S_\mathrm{D2D} \leq \frac{W_p - G_p}{E_p}.
\end{align} 
Note that $W_p$ and $E_p$ depend on the SF used for the preceding downlink frame and the current D2D window, respectively (see~\eqref{eq:Wp} and~~\eqref{eq:Ep}). Thus, the maximum possible value of $S_{\mathrm{D2D}}$ given by~\eqref{eq:SD2D_bound} can vary over the course of the FUOTA session if the SFs change over time (e.g., as in the downlink-SF-selection scheme of~\cite{Bor25}). Our scheme employs a design parameter $S^*$, defined as the largest permitted value of $S_{\mathrm{D2D}}$. The value of $S_{\mathrm{D2D}}$ for a given D2D window is calculated as
\begin{align}
    S_{\mathrm{D2D}} = \min \left\{ \frac{W_p - G_p}{E_p}, S^*\right\}.
\end{align}

\subsection{D2D Transmission Schedule}
Each D2D frame is sent over a randomly chosen superslot within the D2D window. An ED is allowed to send only one D2D frame per D2D window. Furthermore, an ED transmits its D2D frames following consecutive downlink frames. For example, suppose an ED successfully reconstructs the update upon successfully receiving a downlink frame during the \mbox{$J$-th} downlink superslot. Further suppose that the ED decides (using the procedure in Section~\ref{sec:N2D}) to send $N_\mathrm{D2D}$ frames. Then, it sends its first D2D frame in a superslot selected uniformly at random within the D2D window that follows downlink superslot $J \plus \lambda$, where $\lambda$ is a positive integer. The next D2D frame is sent in a randomly chosen superslot following downlink superslot $J \plus \lambda \plus 1$, and so on, until the $N_\mathrm{D2D}$-th frame is sent following downlink superslot $J \plus \lambda \plus N_\mathrm{D2D} \minus 1$.  The integer $\lambda$ accounts for the processing delay incurred by the ED while decoding the update and producing new coded fragments. Its value may vary from ED to ED. 

\subsection{Adaptive Selection of the Number of D2D Frames}
\label{sec:N2D}
The number of D2D frames sent by an ED is determined by the ED's perception of the decoding status of its neighbors. The guiding principle is that an ED should send more D2D frames when few neighbors have received the update, and fewer frames otherwise. To this end, each ED maintains a count of the number of senders from which it received D2D frames prior to its own decoding success. Let $\beta$ denote this count. The ED then computes the \textit{success ratio} $\eta$ as 
\begin{align} \label{eq:succ_ratio}
    \eta = \frac{\beta}{cN_{\mathrm{ED}}},
\end{align}
where $N_{\mathrm{ED}}$ is the total number of recipients (assumed to be announced by the server during the multicast session setup) and the scaling factor $c$ is a constant in the range $(0,1]$. Observe that for $c \equals 1$ and lossless D2D links, $\eta$ equals the fraction of recipients that have already received the update. In practice, an ED is unlikely to receive from all D2D senders, since some of them may be far away, and some frames may be lost due to collisions, fading, and shadowing. Therefore, the denominator is scaled by the constant $c$ in calculating the success ratio. The number of D2D frames to be sent by the ED is then computed as 
\begin{align}
    N_{\mathrm{D2D}} = \max\left\{ (1 - \eta)N^{(\max)}_{\mathrm{D2D}}, N^{(\min)}_{\mathrm{D2D}} \right\}
\end{align}
where $N^{(\max)}_{\mathrm{D2D}}$ and  $N^{(\min)}_{\mathrm{D2D}}$ are design parameters denoting the maximum and minimum number of D2D frames that an ED is permitted to send.  Their values are configured by the server according to the energy profile of the
EDs. In general, $N^{(\max)}_{\mathrm{D2D}}$ and  $N^{(\min)}_{\mathrm{D2D}}$ are small if the EDs have low-capacity batteries.

\subsection{Coded Fragment Generation for D2D Transmissions} 
To generate coded fragments for D2D transmissions, an ED applies the rateless code to the recovered source fragments. To prevent transmission of the same coded fragment by multiple devices, each ED chooses the sequence numbers for its transmitted coded fragments from a unique set of $N^{(\max)}_\mathrm{D2D}$ sequence numbers. Specifically, the $j$th coded fragment from ED $i$ has the sequence number 
\begin{align}
s_j^{(i)} = M + iN^{(\max)}_\mathrm{D2D} + j,  
\end{align}
where $i \equals 0, 1, \ldots, N_{\mathrm{ED}} \minus 1$ and $j \equals 0, 1, \ldots, N^{(\max)}_{\mathrm{D2D}} \minus 1$.  

Note that this approach caps the maximum number of supported EDs per multicast session at 
\begin{align}
    N^{(\max)}_{\mathrm{ED}} = \frac{M_0 - M}{N^{(\max)}_{\mathrm{D2D}}},
\end{align}
where $M_0$ is the largest possible sequence number of any coded fragment. Assuming a 16-bit representation of the sequence number (e.g., as in the 3GPP raptor code), we have $M_0 \equals 65,536$. As a representative example, with $M \equals 10,000$ and $N^{(\max)}_{\mathrm{D2D}} \equals 25$, we obtain $N^{(\max)}_{\mathrm{ED}} = 2214$.

The generation of coded fragments introduces a modest computational cost. 
Owing to the near-linear encoding complexity of Raptor codes, producing $N$ coded fragments requires $\mathcal{O}(N)$ XOR operations at the symbol level. 
Since $N \leq N_{\mathrm{D2D}}^{(\max)}$ in the proposed scheme and $N_{\mathrm{D2D}}^{(\max)}$ is small, the resulting encoding overhead is minimal.

\subsection{Reception at the EDs} 
To receive the gateway's transmissions, the EDs open their receive window during each downlink superslot. Until the gateway completes its first $k$ transmissions, there is no possibility of D2D transmissions. Therefore, the EDs do not attempt receptions over D2D windows during this time. From the $k$-th downlink transmission onward, EDs remain in the receive mode over D2D windows to collect coded fragments. Once the ED has accumulated enough coded fragments to reconstruct the update, it stops listening for FUOTA frames.

\subsection{Remarks on Duty-Cycle Compliance at the EDs} 
Note that the transmission schedule is driven by duty-cycle considerations at the gateway. According to~\eqref{eq:Wp}, from the start of one downlink frame to the start of the next, the fraction of time the gateway transmits does not exceed the maximum allowed duty-cycle. However, the same may not apply to an ED's transmissions. For example, consider a situation in which the gateway uses a smaller SF (and hence transmits shorter frames) than an ED that is broadcasting D2D frames. Since the ED transmits frames with the same periodicity as the gateway, its duty-cycle computed over the $N_{\mathrm{D2D}}$ transmissions will exceed the regulatory limit. But since each ED transmits only a few frames, the long-term duty-cycle (e.g., computed over a one-hour interval) remains well within prescribed limits.

\subsection{Remarks on Synchronization Requirements} 
The FUOTA specifications address device synchronization on the downlink~\cite{FUOTA_SYNC}. The same can be utilized for gateway-to-ED communications in the proposed method. The proposed scheme has an additional synchronization requirement for D2D transmissions. However, this is easily handled by noting that each downlink frame serves as a synchronization beacon for the upcoming D2D window. Thus, no additional control signaling is needed.

\section{Extension to Batched Rateless Coding}
\label{sec:batched}
For large update sizes, it may be desirable to divide the update into multiple batches, each containing a fixed number of source packets, and then apply rateless coding to each batch separately. This helps restrict the computational complexity of decoding. In this section, we generalize our proposed method to handle batched coding. 

Suppose that the update is divided into $\nu$ batches. In our method, the gateway cycles through the batches. The first downlink transmission is from batch 1, the second from batch 2, and so forth, until one frame from each batch has been transmitted. Thereafter, the gateway returns to batch 1 and repeats the process. Each ED maintains the variables $\beta_1, \beta_2, \ldots, \beta_{\nu}$, where $\beta_i$ is the number of senders from which the ED has received D2D frames generated from batch-$i$ (analogous to $\beta$ of Section~\ref{sec:N2D}.) Once the ED decodes batch $i$, it computes the success ratio as $\eta_i \equals \beta_i/(cN_{\mathrm{ED}})$. The number of D2D frames from batch $i$ to be sent by the ED is then calculated as $N_{\mathrm{D2D}}^{(i)} \equals \max\{ (1 \minus \eta_i)\hat{N}^{(\max)}_{\mathrm{D2D}}, \hat{N}^{(\min)}_{\mathrm{D2D}} \}$, where  $\hat{N}^{(\max)}_{\mathrm{D2D}}$ and  $\hat{N}^{(\min)}_{\mathrm{D2D}}$ are the maximum and minimum allowed D2D transmissions per ED per batch, respectively.      

Similar to the scheme of Section~\ref{sec:proposed_scheme}, a D2D window follows a downlink transmission. The number of superslots in the window is determined as described in Section~\ref{sec:D2D_window}. However, a key difference is that each D2D window is reserved for a specific batch. Specifically, a D2D frame from given batch can be sent only within a D2D window that follows a downlink transmission from the same batch. This allows receiving EDs to know which batch is expected during a D2D window; consequently, an ED keeps its receiver on only during those D2D windows that correspond to batches that the ED has not yet decoded, thereby saving energy. To transmit a D2D frame, the sender chooses a superslot at random within an eligible D2D window.

\section{Performance Evaluation}
\label{sec:perf_eval}

Our primary performance metric is the \textit{completion time} at a recipient, since the goal is to reduce update-delivery delay. It is defined as the time since the start of the FUOTA session until the recipient obtains the complete update. We also evaluate the \textit{device activity time}, defined as total duration for which a device is either transmitting, receiving, or scanning for FUOTA frames. Thus, activity time is an indicator of the energy consumption of a device. The simulations are performed in MATLAB. Each data point is obtained by averaging results from 100 simulation runs. 

As in~\cite{Bor25}, we consider the delivery of an update of size 10 kB, divided into 200 fragments of 50 bytes each. The recipients are uniformly distributed over a circle of radius $\mathcal{R}$. The gateway is located at the center. Unless stated otherwise, we consider $400$ recipients, $\mathcal{R} \equals 1$ km,  and the update is not divided into batches. The default values of protocol parameters are $\SF_{\mathrm{D2D}} \equals 10$, $S^* \equals 20$, ${N}^{(\max)}_{\mathrm{D2D}} \equals 25$, ${N}^{(\min)}_{\mathrm{D2D}} \equals 10$, and $c \equals 0.25$. On the downlink, we use  $L \equals 7$, and $w \equals 300$.

The duty-cycle limit is 1\,\%.  The  links have path loss exponent 2.5 and are subject to independently and identically distributed block Rayleigh fading.  The received power at distance $d$ is  \mbox{$R \equals\gamma_0 p_t A d^{-\alpha}$}, where $p_t$ is the transmit power, $A$ is the fading coefficient with pdf $f_A(a) \equals \e^{-a}$ for $a \gteq 0$, and $\gamma_0$ is a constant that depends on antenna gains and transmission wavelength~\cite{Bor25}.

The transmissions experience interference from nodes distributed according to a Poisson point process of intensity $1\!\times\! 10^{-5}$. Each interferer makes pure ALOHA transmissions of frames according to a Poisson arrival process at a rate of 1 frame per 10 minutes. An interfering frame is equally likely to use any of the SFs between 7 and 12. The number of bytes in the payload of an interfering frame takes integer values between 1 and 20, all values being equiprobable.  Each transmission has power 14 dBm, bandwidth 125 kHz, and uses one among 8 orthogonal channels at random.  

For modeling collision-related losses, the dominant-interferer model offers a reliable approximation~\cite{GeR17}, and is employed here. In this model, a frame using SF $i$ and received with power $R$ will be lost due to a collision with another frame using SF $i'$ and power $R'$ if $R/R' \lthan \xi_{i,i'}$, where $\xi_{i,i'}$ denotes the capture threshold. The capture threshold values obtained from~\cite{MSG19} are used in our simulations. The receiver sensitivities used in the simulation are as given in~\cite{GeR17}. 

For rateless coding, we assume the raptor code of~\cite{LGS07}. As shown in~\cite{Bor25}, the probability that an ED fails to decode the update after receiving exactly $l$ coded fragments is $1$ for $l \lthan k$, 0.85 for $l \equals k$, and 0.567 for $l \gthan k$.

\subsection{Benchmark Schemes}
\label{sec:benchmarks}
The following schemes are used as benchmarks in our performance evaluations.

\paragraph{Fixed-SF (FSF)} Among existing FUOTA schemes (none of which employ D2D communications), we seek to identify the one that completes the update delivery session in the shortest time and use it as a benchmark. Since an ED at the cell edge (i.e., located at the perimeter of the circle over which the update recipients are distributed) experiences the maximum delivery delay on average,  the scheme that minimizes delivery delay at an edge ED also minimizes the overall average session completion time. As shown in~\cite{Bor25},  minimum delivery delay at the cell edge can be achieved by using a scheme that transmits rateless-coded frames with a fixed and suitably chosen SF. This is essentially the \textit{Fixed-SF (FSF)} benchmark of~\cite{Bor25}. FSF is based on the method of~\cite{SNY23,SNY24}. We use FSF-$i$ to denote an FSF scheme that uses SF $i$.

\paragraph{Groupless Multi-SF (GL-MSF)} This is the scheme of~\cite{Bor25}. The gateway starts transmissions with SF $L$ \mbox{($7 \leq L \leq 12$)}, and increments the SF by one after every $w$ transmissions. To distinguish it from the group-based scheme of~\cite{MZK23}, we call it the \textit{groupless multi-SF (GL-MSF)} scheme.  

\paragraph{D2D With Perfect Schedule Information (D2D-PSI)} This is an idealization of our proposed scheme. It is identical to the proposed method, except that every recipient knows the exact superslots in which decoded recipients send their D2D frames. This enables a recipient to turn on its receiver only for such slots. Thus, wasteful energy expenditure associated with having to listen over unused D2D superslots (as in the proposed scheme) can be avoided.

\subsection{Numerical Results}
\label{sec:res}

\begin{figure}
\begin{subfigure}{0.5\textwidth}
  \centering
  \includegraphics[scale=0.55]{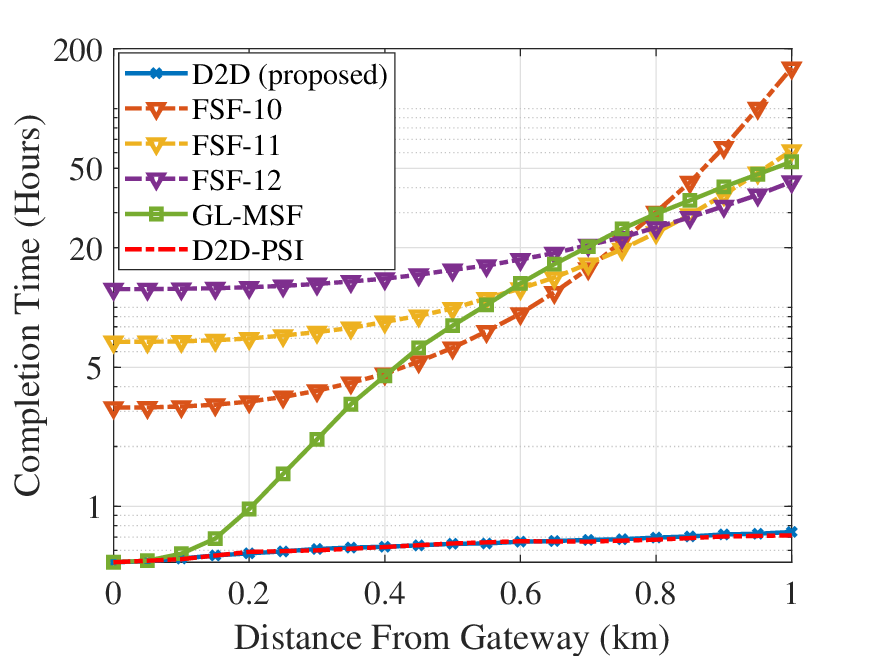}  
  \vspace{-1mm}
  \caption{Average completion time.}
  \label{fig:latency_vs_dist}
\end{subfigure}
\begin{subfigure}{0.5\textwidth}
  \centering
  \includegraphics[scale=0.55]{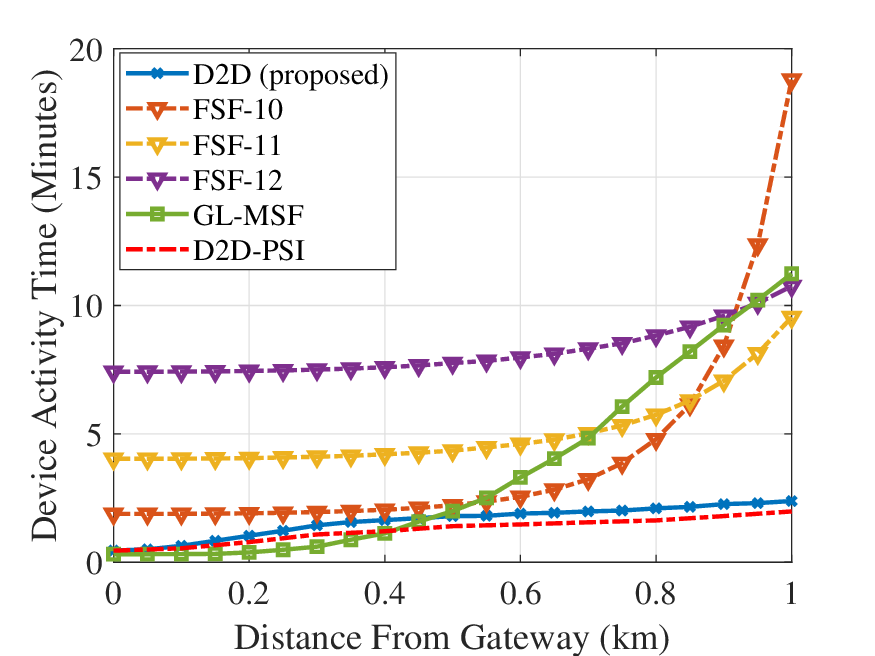}  
  \vspace{-1mm}
  \caption{Average device activity time.}
  \label{fig:energy_vs_dist}
\end{subfigure}
\caption{Impact of node distance.}
\label{fig:vs_Distance}
\end{figure}

\begin{figure}
\begin{subfigure}{0.5\textwidth}
  \centering
  \includegraphics[scale=0.55]{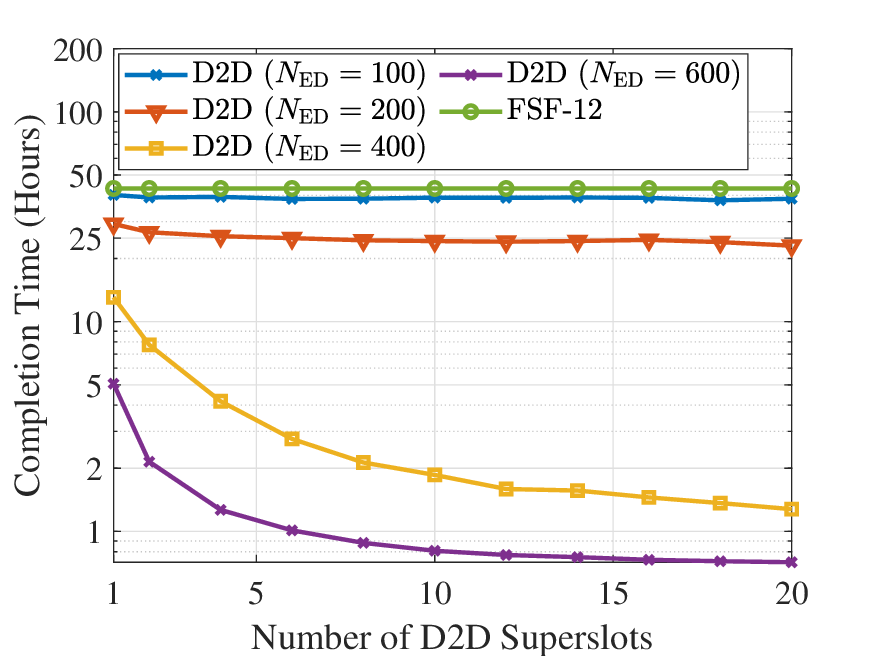}  
  \vspace{-1mm}
  \caption{Average completion time.}
  \label{fig:latency_vs_d2dslots}
\end{subfigure}
\begin{subfigure}{0.5\textwidth}
  \centering
  \includegraphics[scale=0.55]{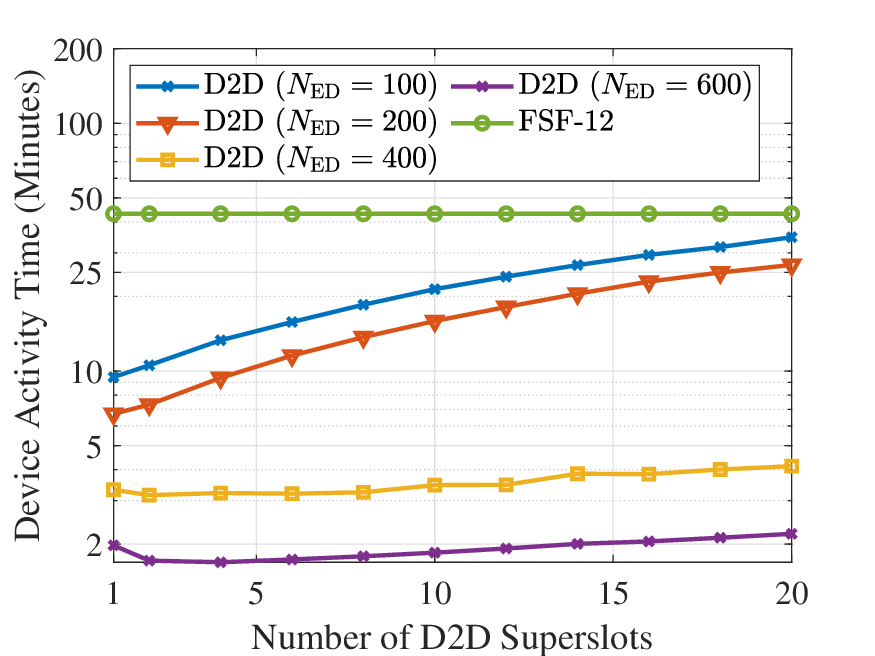}  
  \vspace{-1mm}
  \caption{Average device activity time.}
  \label{fig:energy_vs_d2dslots}
\end{subfigure}
\caption{Impact of recipient count ($N_{\mathrm{ED}}$) and maximum permitted number of superslots per D2D window ($S^*$) on the performance of a device at the cell edge.}
\label{fig:vs_d2dslots}
\end{figure}

Fig.~\ref{fig:latency_vs_dist} plots the average completion time at an ED against the ED's distance from the gateway.  For each scheme, the completion time increases with node distance due to more frequent frame erasures. Among the non-D2D benchmarks (FSF, GL-MSF), FSF-12 minimizes the worst-case completion time at the cost of significantly higher delays at nearby recipients. GL-MSF performs well for nearby recipients, but its worst-case delay slightly exceeds FSF-12. In contrast, our proposed scheme provides low delay at all distances by utilizing D2D links. Its completion time at the cell edge is only 45 minutes, compared to 42 hours for FSF-12. The hypothetical D2D-PSI scheme also provides similar completion times.  This is expected, since the only difference between D2D-PSI and our D2D scheme is that the former permits an ED to sleep during D2D superslots in which no transmissions occur. Such superslots do not contribute to the completion time anyway.   

Fig.~\ref{fig:energy_vs_dist} shows that in the proposed method, EDs near the gateway incur slightly higher device activity time than GL-MSF. This is due to the D2D transmission and reception overhead. For distant recipients, our proposed scheme exhibits much lower activity times than all benchmarks (e.g., at least 75\% reduction at 1 km distance.) Without D2D, distant EDs spend a large fraction of their time on failed reception attempts. With D2D, such attempts are reduced, thus decreasing the overall device activity time despite the D2D overhead.

\begin{figure}
\begin{subfigure}{0.5\textwidth}
  \centering
  \includegraphics[scale=0.55]{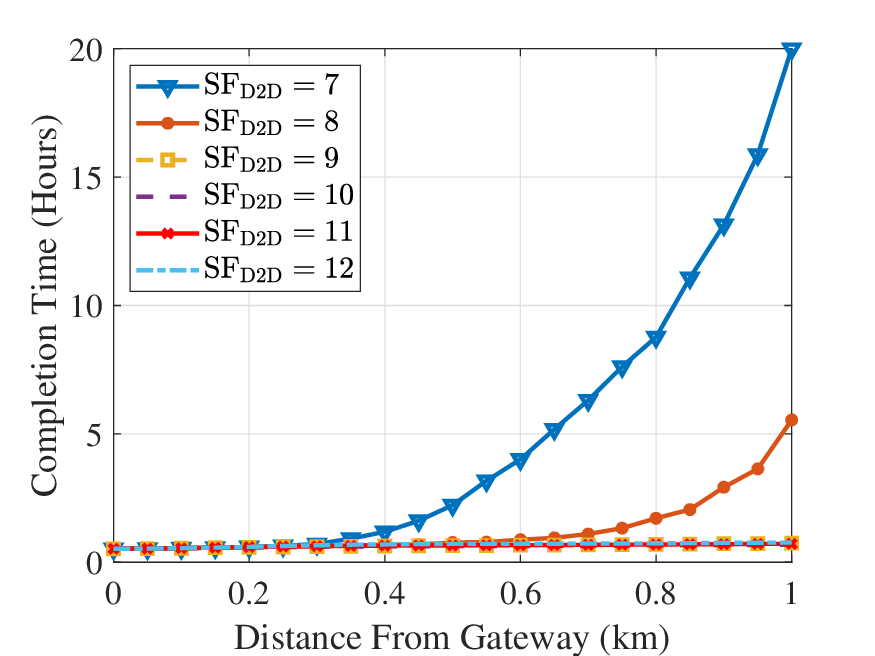}  
  \vspace{-1mm}
  \caption{Average completion time.}
  \label{fig:latency_vs_d2dSF}
\end{subfigure}
\begin{subfigure}{0.5\textwidth}
  \centering
  \includegraphics[scale=0.55]{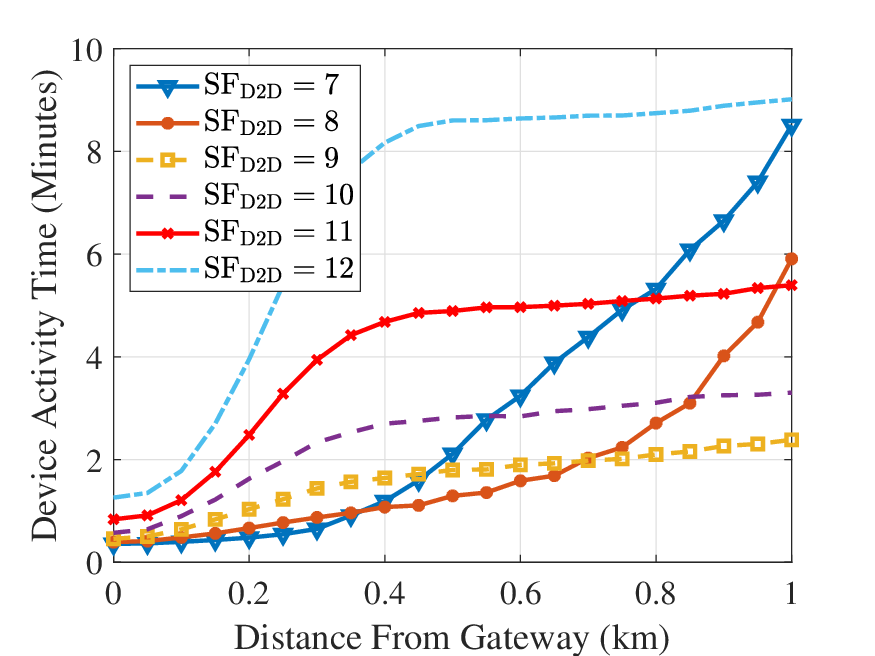}  
  \vspace{-1mm}
  \caption{Average device activity time.}
  \label{fig:energy_vs_d2dSF}
\end{subfigure}
\caption{Impact of the SF used on D2D links.}
\label{fig:impact_D2D_SF}
\end{figure}

Fig.~\ref{fig:latency_vs_d2dslots} illustrates the impact of $S^*$, the maximum permitted number of superslots in a D2D window, on the completion time at the cell edge. We observe that increasing $S^*$ reduces the completion time. Recall that all D2D transmissions scheduled after a given downlink frame must contend for the same $S_{\mathrm{D2D}}$ slots. Increasing $S_{\mathrm{D2D}}$ (by having a larger $S^*$) reduces collisions and improves D2D receptions, expediting update delivery.   However, even with $S^* \equals 1$, substantial reductions in the completion time are observed for 200 or more EDs. In general, having more EDs is beneficial, as more D2D transmissions lead to reduced completion time.

Fig.~\ref{fig:energy_vs_d2dslots}  shows how $S^*$ impacts the device activity time. For a fixed number of EDs, activity time increases with $S^*$, because a recipient spends more time scanning for and/or receiving D2D frames. The only exception is $S^*\equals 1$ with 600 EDs. In this case, many EDs often transmit over the single D2D slot, causing frame loss. Consequently, the system does not benefit from D2D transmissions to the extent that $S^* \gteq 2$ does. We further notice that for a fixed $S^*$, activity time decreases with the ED count. More EDs lead to more D2D frames, permitting unfinished recipients to collect more fragments per D2D window. Consequently, the recipient encounters fewer D2D windows until it receives the complete update.

Fig.~\ref{fig:latency_vs_d2dSF} illustrates how the SF used for D2D frames affects the completion time. Increasing $\mathrm{SF}_\mathrm{D2D}$ reduces the completion time, with large gains in going from SF~7 to SF~9 and only marginal changes thereafter. The corresponding device activity times (Fig.~\ref{fig:energy_vs_d2dSF}) exhibit more complex behavior. For recipients near the gateway, larger $\mathrm{SF}_\mathrm{D2D}$ increases activity time because these nodes rely little on D2D reception (owing to strong direct links from the gateway) and the added transceiver activity is mainly from longer D2D transmissions as $\mathrm{SF}_\mathrm{D2D}$ increases. For distant recipients, device activity is dominated by D2D reception. With small $\mathrm{SF}_\mathrm{D2D}$, these nodes make many unsuccessful reception attempts, increasing activity time without aiding update completion. With large $\mathrm{SF}_\mathrm{D2D}$, reception times are longer due to longer frame durations, but these receptions are more likely to succeed. Thus, increasing $\mathrm{SF}_\mathrm{D2D}$ from 7 onward first reduces device activity time before it increases again. We find SF~9 to provide the lowest activity time at the cell edge for the network parameters considered.

\begin{figure}
    \centering
    \includegraphics[scale=0.55]{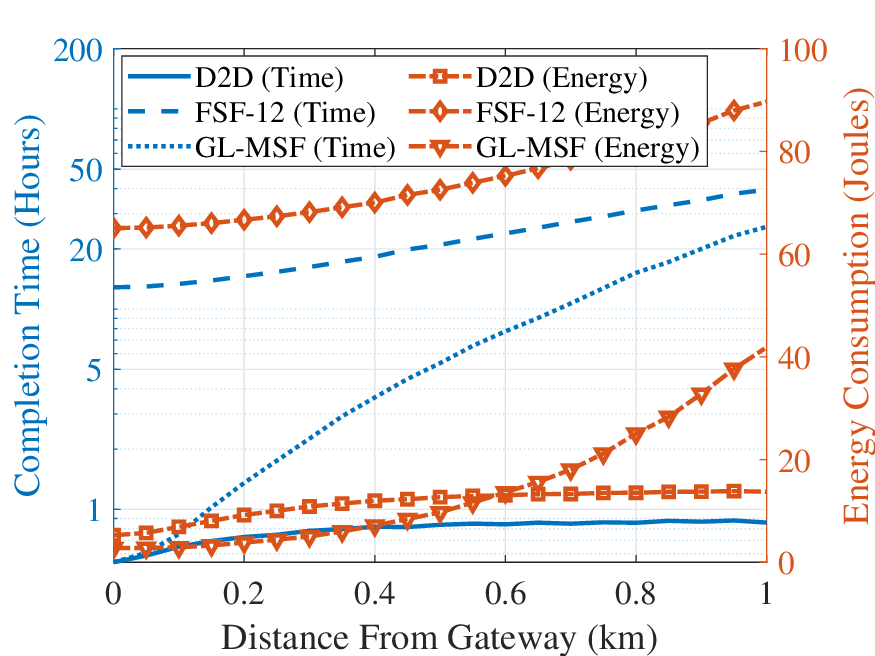}
    \caption{Performance for batched delivery.}
\label{fig:delay_energy_vs_d0_batched}
\end{figure}

Fig.~\ref{fig:delay_energy_vs_d0_batched} shows results for a scenario in which the update is divided into five batches of 2 kB each. To model potential link blockages, we implement log-normal shadowing with a standard deviation of 8 dB on the links. The protocol parameters are $\hat{N}^{(\max)}_{\mathrm{D2D}} \equals 5$, $\hat{N}^{(\min)}_{\mathrm{D2D}} \equals 2$, and $c \equals 0.25$. As with unbatched delivery, the D2D-aided scheme significantly reduces the completion time. Fig.~\ref{fig:delay_energy_vs_d0_batched} also shows the approximate energy consumption by an ED, computed as 
\begin{equation} \label{eq:energy}
E = (I_t \overline{T}_t + I_r \overline{T}_r) V_b,    
\end{equation}
where $I_t$ and $I_r$ are the ED's current consumption while transmitting and receiving, respectively, $\overline{T}_t$ and $\overline{T}_r$ are the average durations for which the ED is in the transmit and receive modes, respectively, and $V_b$ is the battery voltage. To evaluate~\eqref{eq:energy}, we use $I_t \equals 83$~mA and $I_r \equals 38$~mA~\cite{CMV17}, and assume $V_b \equals 3.7$~V. The energy performance of the schemes exhibit trends similar to those for the device activity times of unbatched delivery. The D2D scheme provides large energy savings over FSF, and improves the worst-case energy consumption compared to GL-MSF at the expense of some additional energy consumption for EDs closer to the gateway. 

Table~\ref{tab:energy} shows a break-up of the energy consumption by an ED at the cell edge.  We consider only the energy consumed for FUOTA-related transmissions and receptions. Energy spent in sleep mode is excluded, as it is incurred by all LoRaWAN devices irrespective of participation in a FUOTA session.
As seen from Table~\ref{tab:energy}, FSF and GL-MSF do not require any transmission energy. In contrast, the proposed scheme spends some energy on transmissions, but greatly reduces the reception energy as well as the overall energy consumption. 

\begin{table}

\centering
    \caption{Energy consumption (in Joules) at an ED located at the cell edge}
    \begin{tabular}{c|c|c|c }
         & D2D & FSF-12 & GL-MSF\\ \hline
        Transmission & 2.3 & 0 & 0\\
        Reception & 11.4 & 89.7 & 41.9\\
        Total & 13.7 & 89.7 & 41.9\\
    \end{tabular}
    \label{tab:energy}
\end{table}

Overall, our numerical evaluations demonstrate that the proposed method yields consistent latency and energy improvements across a wide range of operating conditions. At the cell edge of dense LoRaWAN deployments, orders-of-magnitude reductions in update latency and multi-fold reductions in energy consumption are observed. We have observed that the performance gains decrease only in very sparse networks (e.g., fewer than 30 devices per square kilometer; see Fig.~\ref{fig:vs_d2dslots}) or when D2D connectivity is severely constrained by strong shadowing or fully obstructed links. Although not numerically illustrated here, our design has inherent adaptability to mobility, as D2D communications in our method require no topology information and enable opportunistic reception by nearby EDs.

\section{Conclusion}
This work presents a D2D-aided mechanism that substantially accelerates over-the-air update delivery in large-scale LoRaWANs without requiring additional infrastructure or feedback signaling. By reducing delivery delays with the help of cooperative D2D transmissions, the proposed approach enables timely dissemination of urgent security updates and improves the efficiency and convergence of iterative distributed learning, including federated learning. As such, it provides a practical foundation for secure and intelligent large-scale industrial and agricultural IoT deployments while preserving the low-power and wide-area benefits of LoRaWAN.

\bibliographystyle{IEEEtran}
\bibliography{refs.bib}

\end{document}